\documentclass[pra, twocolumn]{revtex4}

\usepackage{amsmath}
\usepackage{amsfonts}
\usepackage{amssymb}
\usepackage{amsbsy}

\newcommand{\ket}[1]{|#1\rangle}
\newcommand{\bra}[1]{\langle#1|}
\newcommand{\braket}[2]{\langle #1|#2\rangle}
\newcommand{\proj}[1]{|#1\rangle\langle#1|}
\newcommand{\bx}{\mathbf{x}}
\newcommand{\by}{\mathbf{y}}
\newcommand{\balpha}{\boldsymbol{\alpha}}

\begin{document}

\title{Information causality from an entropic and a probabilistic perspective}

\author{Sabri W. Al-Safi} \email{S.W.Al-Safi@damtp.cam.ac.uk}
\affiliation{DAMTP, Centre for Mathematical Sciences, Wilberforce Road, Cambridge CB3 0WA, UK}

\author{Anthony J. Short}\email{A.J.short@damtp.cam.ac.uk}
\affiliation{DAMTP, Centre for Mathematical Sciences, Wilberforce Road, Cambridge CB3 0WA, UK}

\begin{abstract}
The information causality principle is a generalisation of the
no-signalling principle which implies some of the known restrictions on
quantum correlations. But despite its clear physical motivation, information causality is
formulated in terms of a rather specialised game and
figure of merit. We explore different perspectives on information
causality, discussing the probability of success as the figure
of merit, a relation between information causality and the non-local `inner-product game',
and the derivation of a quadratic bound for these games. We then examine an entropic
formulation of information causality with which one can
obtain the same results, arguably in a simpler fashion.
\end{abstract}

\maketitle

\section{Introduction}

Quantum theory has many strange properties, but perhaps the most surprising is that of \emph{non-locality}. Some quantum states, known as \emph{entangled} states, cannot be described by giving a separate quantum state for each system, or even by a probabilistic mixture of such states. This is not just an artifact of the mathematical formalism; many entangled states give rise to observable correlations which cannot be explained by any local model \cite{bell65, chsh69, aspect82}. However, an important caveat is that these non-local correlations cannot be used for superluminal signalling.

Although this area has been extensively studied, we still don't have a good intuition about which non-local correlations are achievable in quantum theory, and what they can be used for. They are certainly helpful in some non-local tasks \cite{nielsen&chuang, cleve04}, but it has been shown that even stronger correlations are possible without generating superluminal signalling \cite{pr94}. Furthermore, there have recently been a number of results describing non-local tasks for which quantum entanglement is not helpful at all, whilst stronger non-local correlations give an advantage \cite{linden06, pawlowski09, almeida10}. By gaining a better understanding of quantum non-locality, we hope to hone our intuitions about its information-theoretic uses, and perhaps learn more about why nature is quantum.

In this paper, we will discuss one particular non-local task for which quantum non-locality is not helpful (at least with the original figure of merit), known as \emph{information causality} \cite{pawlowski09}. This is an appealing principle which one would reasonably expect to hold, and which quantum theory obeys, yet which can be violated using correlations slightly stronger than quantum theory permits \cite{pawlowski09, pawlowski09.2}.

 Information causality relates to a particular type of game: a bit string $\bx$ of length $n$ is chosen uniformly at random and given to Alice, whilst Bob is given a random number $k$, $(1 \leq k \leq n)$. Alice may then send an $m$-bit message $\balpha$ to Bob, after which Bob must try to guess $x_k$, the $k$'th bit of Alice's original bit-string. Bob's guess when his input is $k$ is denoted $\beta_k$. The parties may decide on a joint strategy and may initially share correlated resources but play from separate locations.

The information causality principle states that
\begin{equation} \label{iceqn}
I \equiv \sum_{k=1}^n I_c(x_k : \beta_k ) \leq m,
\end{equation}
where $I_c (X : Y )$ denotes the classical mutual information of variables $X$ and $Y$ \footnote{$I_c (X : Y )= H(X) + H(Y) - H(XY)$, where $H(X)$ is the classical Shannon entropy $H(\{p_i\}) =-\sum_i p_i \log_2 p_i$. }. The intuition behind this bound is that the total information that Bob can access about Alice's bits cannot exceed the size of the message she sent. Indeed, the inequality in (\ref{iceqn}) is saturated if Alice simply sends to Bob the first $m$ bits of $\bx$, so that $I_c(x_k : \beta_k) = 1$ if $1 \leq k\leq m$, and $0$ otherwise.

It is proven in \cite{pawlowski09} that information causality is obeyed in both the quantum and classical world. However, it can be violated in worlds governed by different physical laws (such as `box world' \cite{barrett05, short&barrett10}, which permits all non-signalling correlations). In what follows, we first discuss probability of success in the information causality game. We then derive a bound which relates information causality to a different non-local game, in which Alice and Bob must compute the inner product of two bit strings. Finally, we will explore an alternative formulation and derivation of information causality based on entropy rather than mutual information.

\section{Probability of success for information causality }

Although quantum entanglement gives no advantage over a classical strategy in the information causality game when $I$ (defined by \eqref{iceqn}) is the figure of merit, it is not true that every quantum strategy can be classically simulated. In fact, if probability of success is used as the figure of merit instead, it can easily be seen that entangled quantum states allow one to do better than in the classical world. For example, in a simple version of the game in which $n=2$ and $m=1$, the optimal classical probability of success is $\frac{3}{4}$ (e.g. when Alice sends Bob $\alpha=x_1$ and he guesses $\beta_k=\alpha$, they always win when $k=1$ and win half the time when $k=2$). However, by exploiting well-known quantum violations of Bell inequalities, Alice and Bob can achieve a success probability of $\frac{2+\sqrt{2}}{4}$. To do this, Alice and Bob first generate bits $a$ and $b$ satisfying $a \oplus b = (x_1 \oplus x_2)(k-1)$  with probability $\frac{2+\sqrt{2}}{4}$, where $\oplus$ denotes addition modulo 2. This is equivalent to the quantum Tsirelson bound for the CHSH inequality  \cite{chsh69, tsirelson80}. Then Alice sends Bob $\alpha=a \oplus x_1$ and Bob outputs $\beta_k = b \oplus \alpha$ \cite{vandam05, pawlowski09}.

It is also possible to obtain very different values of $I$ for strategies with the same probabilities of success. As above, Alice can send Bob her first bit to obtain $I = 1$ and  probability of success $\frac{3}{4}$; alternatively, Alice and Bob can randomly ``mix'' this strategy with one where Alice sends Bob her second bit and he outputs it, so that the overall probability of success is the same but
\begin{align}
I &= I_c(x_1 : \beta_1)+ I_c(x_2 : \beta_2) \nonumber \\
&=2\left( 1 - H\left(\left\{\frac{3}{4},\frac{1}{4}\right\}\right)\right) \nonumber \\
&\approx 0.38.
\end{align}
Furthermore, it is clear that a small amount of noise added to the first strategy will do better than the second strategy in terms of $I$, but worse in terms of success probability, so these two figures of merit are not monotonically related.

The optimal classical strategy to maximize the probability of success in the case when $m=1$ has already been derived for general $n$, in the context of Random-Access Coding \cite{ambainis08}. It is attained by using the ``majority-vote'' strategy, in which Alice simply sends Bob the bit that most frequently occurs in her string. This gives success probability
\begin{equation} \label{classicalsuccess}
 P^C_{success} = \frac{1}{2}\left(1 + \frac{1}{2^{n-1}}\binom{n-1}{\lfloor\frac{n-1}{2}\rfloor}\right)
\end{equation}

Using Stirling's approximation, one can derive the asymptotic behaviour of this probability:
\begin{equation}
 P^C_{success} \approx \frac{1}{2}\left(1 + \sqrt{\frac{2}{\pi n}}\right)
\end{equation}

We show in the next section that the optimal  quantum probability of success for the same situation is
\begin{equation} \label{quantumsuccess}
 P^Q_{success} = \frac{1}{2}\left(1 + \sqrt{\frac{1}{ n}}\right),
\end{equation}
which is always strictly larger than the classical limit. This extends a result obtained in \cite{pawlowski10} for particular $n$ \footnote{Specifically for $n=2^k 3^j$ where $j$ and $k$ are integers.}.

Interestingly, (\ref{quantumsuccess}) is also the optimal success probability when Alice is allowed to send a qubit to Bob instead of a classical bit, but Alice and Bob do not share an entangled state \cite{ambainis08, pawlowski10}.

The probability of success has a clean operational interpretation as a figure of merit: it is the asymptotic fraction of games one would expect to win over many independent repetitions. Although it sounds appealing, the operational meaning of $I$ is less natural. In particular, suppose Alice and Bob play the game many times, then Alice is told Bob's input $k$ for each round, and she sends him some supplementary classical information which (together with his guesses $\beta_k$) he must use to output the correct value of $x_k$ for each round. The average amount of supplementary information per round which Alice must send Bob is $(1-I/n)$. This follows from a result of \cite{slepian73} that the asymptotic amount of information (using coding over many rounds) required to learn $x_k$ given that you hold $\beta_k$  is $H(x_k|\beta_k)= H(x_k) - I(x_k : \beta_k)$.

However, although $I$ is a less natural \emph{a priori} figure of merit than success probability, its appeal lies in the simplicity of the bound given by (\ref{iceqn}). In particular, the maximum value of $I$ is the same for classical or quantum strategies, and can be simply stated for any message length $m$  (by contrast, the maximum success probabilities given by (\ref{classicalsuccess})-(\ref{quantumsuccess}) are complicated, depend on $n$ and only apply when $m=1$).

\section{Information causality and the inner product game}

Given that the mutual information is a complicated non-linear function of the associated probabilities, it is surprising that the bound given by  information causality can be used to derive the Tsirelson bound, which can be understood as a bound on the quantum success probability for a particular non-local game \footnote{In particular, Alice and Bob are given uniformly random bits $x$ and $y$, and must output bits $a$ and $b$ such that $a\oplus b = xy$. The Tsirelson bound is  $P^Q_{success} \leq \frac{2+\sqrt{2}}{4}$, whilst the CHSH inequality gives $P^C_{success} \leq \frac{3}{4}$}. Even more surprisingly,  information causality can be used to generate part of the curved surface of the set of achievable quantum correlations \cite{pawlowski09.2}.

To investigate this, we note that the proof of the Tsirelson bound given in \cite{pawlowski09} can be decomposed into several steps. The first is to prove that the information causality principle $I \leq m$ implies a bound $\sum_{k=1}^n [1-h(P_k))] \leq m$ on the binary entropy \footnote{$h(P_k) = -P_k \log_2 P_k- (1-P_k) \log_2 (1-P_k)$} of the success probability $P_k$ given a particular input for Bob. This entropic bound can be transformed into a quadratic bound on the \emph{bias} $E_k=(2P_k-1)$ achieved in the game by noting that $1-h(P_k) \geq \frac{E_k^2}{2 \ln 2}$. The information causality principle can therefore be used to generate the bound
\begin{equation} \label{ICquadraticbound}
\sum_{k=1}^n E_k^2 \leq 2 m \ln 2
\end{equation}
Finally, the authors consider a particular strategy for playing the game in which $m=1$ and $n$ is a power of 2, and show that the ability to generate correlations violating the Tsirelson bound would allow one to violate (\ref{ICquadraticbound}) for sufficiently large $n$. Hence, given information causality, the Tsirelson bound holds.

As the quadratic bound given by (\ref{ICquadraticbound}) plays a key role in deriving the Tsirelson bound from information causality, it is interesting to investigate such bounds directly in quantum theory. To facilitate this, we first consider a seemingly unrelated non-local game, in which the aim is to produce the inner product of two bit strings. In this inner product game, Alice and Bob are given uniformly random $n$-bit strings $\bx$ and $\by$ respectively. Then without communicating, Alice and Bob must output bits $a$ and $b$ respectively such that $a \oplus b = \bx \cdot \by$, where $\bx \cdot \by = x_1 y_1 \oplus x_2 y_2 \ldots \oplus x_n y_n$.

The ability to win the inner product game perfectly would allow the parties to non-locally compute any function of their inputs \cite{vandam05}, and therefore to solve any communication complexity problem with only a single bit of communication.

We can derive a bound on the inner product game which is very similar to (\ref{ICquadraticbound}). Assume that Alice and Bob share an initial entangled state $\ket{\psi}$, and their outputs are obtained by measuring the operators $\hat{a}_{\bx}$ and $\hat{b}_{\by}$ respectively  (with eigenvalues 0,1). The bias they achieve in the game when they are given inputs $\bx$ and  $\by$ is
\begin{equation}
E_{\bx \by}  =   \bra{\psi} (-1)^{\hat{a}_{\bx} + \hat{b}_y + \bx \cdot \by }\ket{\psi},
\end{equation}
where
\begin{equation}
P^Q_{success} = \frac{1}{2} \left( 1 +  \frac{1}{2^{2n}} \sum_{\bx \by} E_{\bx \by}  \right)
\end{equation}
Similarly, the average bias they achieve when Bob is given $\by$ and we average over Alice's input is given by $ E_{\by} = \frac{1}{2^n}  \sum_{\bx}  E_{\bx \by}$.

To derive a quadratic bound, we adopt a similar approach to \cite{linden06}. We define the normalised states
\begin{eqnarray}
\ket{A} &=& \frac{1}{\sqrt{2^n}} \sum_{\bx}  (-1)^{\hat{a}_{\bx}} \ket{\psi} \otimes \ket{\bx} \\
\ket{B_{\by}} &=& \frac{1}{\sqrt{2^n}} \sum_{\bx}  (-1)^{\hat{b}_{\by}+\bx \cdot \by}  \ket{\psi} \otimes \ket{\bx}
\end{eqnarray}
where the $\ket{B_{\by}}$ states form an orthonormal set satisfying $\braket{B_{\by}}{B_{\by'}} = \delta{\by \by'}$.

It follows that
\begin{eqnarray}
 \sum_{\by}  E_{\by}^2 &=& \sum_{\by} \braket{A}{B_{\by}}^2 \nonumber \\
 &=& \bra{A} \left( \sum_{\by}  \proj{B_{\by}} \right) \ket{A}\nonumber \\
 &\leq& 1  \label{innerproductbound}
 \end{eqnarray}
 where in the last step we have used the fact that  $\sum_{\by}  \proj{B_{\by}}$ is a projector and $\ket{A}$ is normalised. A similar result was obtained independently by Pawlowski and Winter, using a  different method, and described very recently in \cite{pawlowski11} \footnote{We can also prove an analogue of their more general bound, where Alice's input has probability distribution $p(\bx)$, by replacing $\ket{A}$ with the unnormalized state $\sqrt{2^n} \sum_{\bx} p(\bx)  (-1)^{\hat{a}_{\bx}} \ket{\psi} \otimes \ket{\bx}$. This gives $\sum_{\by}  E_{\by}^2 \leq 2^n \sum_{\bx} p(\bx)^2$.}.

We can also obtain a bound on the probability of success from (\ref{innerproductbound}), by taking
\begin{eqnarray}
P^Q_{success} &=& \frac{1}{2} \left( 1 +  \frac{1}{2^n}  \sum_{ \by}  E_{ \by}  \right) \nonumber \\
&\leq& \frac{1}{2} \left( 1 +  \sqrt{  \frac{1}{2^n} \sum_{\by}  E_{ \by}^2 } \right)\nonumber \\
&\leq& \frac{1}{2}  \left( 1 +   \frac{1}{\sqrt{2^n}} \right) \label{probsuccess}
\end{eqnarray}
When $n=1$, the inner-product game is equivalent to the CHSH game \cite{chsh69}, and this bound on the success probability corresponds to the usual Tsirelson bound.

The bound given by (\ref{innerproductbound}) actually holds regardless of the probability distribution over Bob's input $\by$. This generalisation allows us to derive a bound on a non-local version of the information causality game, in which Alice is given a random $n$-bit string $\bx$, Bob is given a random number $k$ satisfying $1\leq k \leq n$, and they attempt to produce bits $a$ and $b$ such that $a \oplus b = x_k$ without communicating. If Bob's bit-string in the inner-product game is chosen at random from the set of bit-strings containing a single one (i.e. from the bit-strings of Hamming weight 1), with $k$ denoting the position of the non-zero bit in $\by$, then $\bx \cdot \by = x_k$. In this case, the  inner-product game is the same as the non-local information causality game and (\ref{innerproductbound}) gives
 \begin{equation}
  \sum_k E_k^2 \leq 1  \label{ICsimple}.
  \end{equation}
Note that this is a stronger bound than (\ref{ICquadraticbound}), which was obtained from information causality. Similarly, an analogous derivation to (\ref{probsuccess}) gives
\begin{equation} \label{quantumsuccess2}
 P^Q_{success} \leq \frac{1}{2}\left(1 + \sqrt{\frac{1}{ n}}\right),
\end{equation}

We now show that the bound given by (\ref{ICsimple}) can be saturated in quantum theory for any choice of $E_k$. It was proved in \cite{tsirelson87} (and used in \cite{pawlowski11}) that for any set of real vectors $\mathbf{u}_{\bx}$ and  $\mathbf{v}_{\by}$ of at most unit length, we can find a quantum state $\ket{\psi}$ of a bipartite system, and binary-valued operators $\hat{a}_{\bx}$ and $ \hat{b}_{\bx}$ (which can be measured locally on subsystems A and B), such that
\begin{equation}
\bra{\psi} (-1)^{\hat{a}_{\bx}+ \hat{b}_{\bx}}\ket{\psi} = \mathbf{u}_{\bx}^T \mathbf{v}_{\by}.
\end{equation}
and hence
 \begin{equation}
 E_{\bx\by} =  (-1)^{\bx \cdot \by}\mathbf{u}_{\bx}^T \mathbf{v}_{\by}.
 \end{equation}
For any desired biases $E_{\by}$ satisfying $\sum_{\by} E_{\by}^2 \leq 1$ we can consider the vectors
\begin{eqnarray}
 \mathbf{u}_{\bx} &=& \sum_{\by} (-1)^{\bx \cdot \by} E_{\by} e_{\by}, \\
 \mathbf{v}_{\by} &=& e_{\by},
 \end{eqnarray}
 where $e_{\by}$ denotes an orthonormal basis for a real vector space with dimension equal to the number of different inputs for Bob. This gives $E_{\mathbf{xy}}=E_{\by}$, and hence we can achieve any set of biases satisfying (\ref{innerproductbound}).
  In particular, we could obtain an equal bias for all of Bob's possible inputs in the non-local information causality game ($E_k=\frac{1}{\sqrt{n}}$) which would achieve the optimal probability of success $\frac{1}{2} \left(1 + \frac{1}{\sqrt{n}}\right)$ given by (\ref{quantumsuccess2}).

  Although these results apply to the non-local version of the information causality game, any strategy can be transferred to the original version of the game  with $m=1$, with the same probability of success. Alice simply sends the message $\alpha=a$ to Bob, and he outputs $\beta=a \oplus b$. This is not the only type of strategy which is possible in the original information causality game (e.g. Bob's measurement could depend on Alice's message). However, in \cite{pawlowski10} an identical inequality to (\ref{quantumsuccess2}) is derived for the original  game, hence the optimal strategy for the non-local version  of the game is also optimal when transferred to the original game.  Note that the strategies used in \cite{pawlowski09} to derive the Tsirelson bound, and to achieve perfect success given arbitrary non-signalling resources, are also of this form.

  The bound $\sum_{\by} E_{\by}^2 \leq 1$ for the inner product game seems to capture a great deal about the possible quantum correlations, yet note that this inequality can also be saturated by a classical strategy. In particular, if Alice and Bob output $a=\balpha \cdot \bx$ and $b=0$, they will achieve a bias of 1 when $\by=\balpha$ and 0 in every other case.

\section{Information causality from Entropy} \label{entropysec}

Given the above, it appears that the particular mathematical form of the mutual information is not central in defining the boundary of the set of quantum correlations (as the proof proceeds via a quadratic bound), and the choice of $I$ rather than probability of success as the figure of merit seems somewhat arbitrary. However, the fact that quantum theory obeys information causality actually follows from the existence of a natural extension of the classical mutual information to quantum states. Can we focus on this as a defining property of quantum theory?

In general probabilistic theories, the state of a system is characterised by a complete description of the probability of each measurement outcome, for any possible measurement on that system\cite{ hardy01, barrett05, chiribella10}. A specific probabilistic theory is defined by allowing certain types of systems, and certain states on those systems: for example, classical theory consists of systems specified by a single probability distribution (such as a ball in one of several boxes). In any such theory, it was shown in \cite{pawlowski09} that the information causality principle $I \leq m$ will hold if an analogue of the mutual information $I(X:Y)$ can be defined for all systems $X$ and $Y$ (which may be composite) with the following properties:
\begin{description}
 \item[(i) Consistency:] Whenever $X$ and $Y$ are classical systems, $I$ reduces to the classical mutual information, $I(X:Y) = I_c(X:Y)$
 \item[(ii) Data Processing:] Whenever a transformation is performed on $Y$ alone, $\Delta I(X:Y) \leq 0 $
 \item[(iii) Chain Rule:] For all tripartite systems $X$, $Y$, $Z$ \begin{equation*}I(X : YZ) - I(X : Z) = I(XZ : Y) - I(Z : Y)\end{equation*}
 \item[(iv) Symmetry:] $I(X:Y) = I(Y:X)$
 \item[(v) Non-negativity:] $I(X:Y) \geq 0$.
\end{description}
It is well known that all of these properties are satisfied by $I_q$ and $I_c$, the quantum and classical versions of the mutual information. The proof of information causality also assumes the validity of some natural operations, in particular the ability to discard a system, or to prepare a system in a state determined by the value of a classical variable.  These transformations can be defined for any theory in the general probabilistic framework of \cite{barrett05}. If we consider discarding both $X$ and $Y$, we can actually derive non-negativity from the symmetry and data-processing conditions, since (denoting a discarded system by $\varnothing$)  $I(X:Y) \geq I(X:\varnothing) = I(\varnothing : X) \geq  I(\varnothing : \varnothing) =0$, hence condition $(v)$ can easily be eliminated \footnote{Private communication with Marcin Pawlowski}.

 However, while the other properties seem intuitively reasonable, property $(iii)$ seems like a strange demand. Furthermore, the fact that the mutual information necessarily concerns a pair of systems makes it a somewhat complicated quantity.

In the remainder of this section, we show that the Information Causality principle follows more simply from the existence of `good' measure of entropy in a general theory. In particular, the entropy only concerns a single system (although this may be composite), and is only required to obey two conditions.
\begin{description}
 \item[(I) Consistency] If system $X$ is classical, $H(X)$ reduces to the classical entropy, $H(X)=H_c(X)$.
 \item[(II) Evolution with an ancilla] For any two systems $X$ and $Y$, whenever a transformation is performed on $Y$ alone,
\begin{equation}
 \Delta H(XY) \geq \Delta H(Y) \label{evolutionassumption}
\end{equation}
\end{description}
Condition $(I)$ says that $H$ gives the asymptotic compression rate for classical data. Condition $(II)$ can be understood intuitively as saying that a local transformation can generate more uncertainty than its effect on an individual subsystem would suggest, as it can destroy but not create correlations. If we also define a conditional entropy analogously to the quantum and classical quantity, as $H(X|Y) = H(XY) - H(Y)$, we can alternatively re-express (\ref{evolutionassumption})  as $\Delta H(X | Y)  \geq 0$. We can also express $(II)$ symmetrically as the requirement that  $\Delta H(XY) \geq \Delta H(X) + \Delta H(Y)$ under local transformations on $X$ and $Y$.

Given an entropy function obeying the above conditions, we can define a mutual information analogously to the quantum and classical case as
\begin{equation}
I(X:Y) = H(X) + H(Y) - H(XY).
\end{equation}
This automatically ensures that conditions $(iii)$ and $(iv)$ are satisfied, removing the awkwardness of having to postulate the Chain Rule, and $(i)$ and $(ii)$ follow trivially from $(I)$ and $(II)$ respectively.

The existence of an entropy function with properties $(I)$ and $(II)$  is therefore sufficient to derive information causality. Conversely,  in any theory in which one can violate Tsirelson's bound, it must be \emph{impossible} to define an entropy which satisfies assumptions $(I)$ and $(II)$. Several entropies which can be applied to any probabilistic theory, and which always obey $(I)$, have been proposed in \cite{short10, barrett10,kimura10}. A different set of entropic conditions which can be used to derive information causality were discussed in \cite{barrett10}.

It's not hard to deduce some other standard properties of the entropy from conditions $(I)$ and $(II)$:

\begin{description}
\item[Subadditivity] By discarding $Y$, we find from $(II)$ that the entropy is subadditive:
\begin{equation}
H(XY) \leq H(X) + H(Y) \label{subadd}.
\end{equation}
When $X$ and $Y$ are independent systems, we can also prepare $Y$ locally, which implies that $H(X,Y) = H(X) + H(Y)$ in this case.
\item[Strong Subadditivity] By discarding $Z$ from the composite $YZ$ in the tripartite system $XYZ$, we obtain strong subadditivity:
\begin{equation}
H(XYZ) + H(Y) \leq H(XY) + H(YZ).
\end{equation}
This inequality is equivalent to subadditivity of the conditional entropy. It can also be iterated for a larger number of systems to give:
\begin{equation}\label{itercondsubadd}
 H(X_1 \ldots X_n| Y) \leq H(X_1|Y) + \ldots + H(X_n|Y).
\end{equation}

\item[Positivity of Classical Entropy]
Uncertainty about the state of a classical system $X$ can never be negative, even when one conditions on an arbitrary system $Y$.
\begin{equation} \label{poscondclassent}
\text{System X is classical} \Rightarrow H(X|Y) \geq 0
\end{equation}
We argue this last result in the following way: the state of $X$ is described by a probability distribution on a finite set $E$ of outcomes, and for each outcome $e \in E$ there is a corresponding reduced state $\sigma^{Y|e}$ of $Y$. We can therefore obtain the joint state of system $XY$ by a local transformation on $Y$ from a classical system that is initially perfectly correlated with (and identical to) $X$. Before the transformation the conditional entropy is given by $H(X|X)$, and so is non-negative by $(I)$. Then by $(II)$, after the transformation $H(X|Y)$ must also be non-negative.

\end{description}

Information causality can be proven from the existence of an entropy satisfying $(I)$ and $(II)$ by first constructing the mutual information and then applying the proof of \cite{pawlowski09}. However, it can also be proved more directly using the properties of the entropy derived above, and this yields a slight generalisation of the information causality principle.

Bob's guess $\beta_k$ is derived solely from Alice's message $\balpha$ and Bob's system $B$ before that message is sent. Thus whatever the strategy, there is a  transformation from $(\balpha,B)$ to $\beta_k$ for each $k$:
\begin{align}
\sum_k H_c(x_k | \beta_k) &\geq \sum_k H(x_k |\balpha, B) \nonumber \\
&\geq H(\bx | \balpha, B)  \nonumber \\
&= H(\bx, \balpha, B) - H(\balpha,B) \nonumber\\
&\geq H(\bx, \balpha,B) - H(B) - H(\balpha) \nonumber\\
&= H(\bx,\balpha,B) - H(\bx,B) + H(\bx) - H(\balpha) \nonumber\\
&= H(\balpha|\bx,B) + H(\bx) - H(\balpha) \nonumber\\
&\geq H_c(\bx) - H_c(\balpha) \label{condentropyeqn}
\end{align}
This is a generalized form of the information causality principle which makes no assumption on the distribution on Alice's input $\bx$. It can be interpreted as saying that the remaining uncertainty that Bob has about Alice's bits after guessing  must be more than the original uncertainty about her inputs minus the information gained by the message. In the special case in which Alice's inputs are independent, $H_c(\bx)=\sum_k H_c(x_k)$,  and we can rearrange (\ref{condentropyeqn}) to get
\begin{equation}
\sum_k I(x_k : \beta_k) \leq H_c(\balpha) \leq m.
\end{equation}
as in \cite{pawlowski09}.

\section{Conclusions}

Considering probability of success in the information causality game, we see that quantum theory gives an advantage which is not captured by the figure of merit $I$ which is bounded by (\ref{iceqn}). Investigating how these probabilities are involved in deriving Tsirelson's bound from information causality \cite{pawlowski09} leads us to a quadratic quantum bound
\begin{equation}
\sum_{\by} E_{\by}^2 \leq 1
\end{equation}
on the biases achieved given different inputs for Bob in the non-local inner product game. This applies for an arbitrary distribution over Bob's inputs, and hence to the non-local version of the information causality game. This is another example of a bound which quantum and classical correlations can both saturate, but stronger non-local correlations can violate. Furthermore, the fact that quantum correlations allow one to achieve any set of biases satisfying this rule means that it captures a significant amount about the set of quantum correlations. Can we construct useful quadratic bounds on quantum performance in other nonlocal tasks?

Instead of considering information causality as a constraint on possible physical theories, it may be helpful to think of it as a consequence of the existence of a `good' measure of entropy in the theory. Indeed, we have shown that information causality can be derived given any extension of the entropy from classical to more general systems which satisfies $\Delta H(XY) \geq \Delta H(X) + \Delta H(Y)$ under local transformations. Conversely, any theory which violates information causality (such as box world) cannot have an entropy defined in it which obeys the above evolution law and agrees with the Shannon entropy for classical systems.

Given the above results, as well as those of \cite{short10, barrett10, kimura10}, it seems that the existence of a `good' entropy for quantum theory, which shares so many of the properties of the classical entropy, is very special within the class of general probabilistic theories. Are there other theories for which one can define an entropy satisfying $(I)$ and $(II)$, or is this a defining feature  of quantum theory\footnote{Of course, we could consider a restriction of quantum theory which would share the von-Neumann entropy, but the interesting question is to consider theories which cannot be simulated by quantum theory}? The existence of such an entropy potentially places stronger bounds on quantum theory than information causality alone. It would be interesting to look for  other  games where quantum theory can do no better than classical when such an entropy exists.

\vspace{0.3cm}
\emph{Additional note}: Very recently, similar results to those in section \ref{entropysec} have been obtained independently in \cite{dahlsten11}

 \acknowledgments The authors would like to thank Marcin Paw{\l}owski and Sandu Popescu for interesting discussions. AJS also  acknowledges the support of the Royal Society.

\bibliographystyle{ieeetr}

\end{document}